# RenderSelect : a Cloud Broker Framework for Cloud Renderfarm Services

[Pick the date]

[Type the abstract of the document here. The abstract is typically a short summary of the contents of the document. Type the abstract of the document here. The abstract is typically a short summary of the contents of the document.]

# RenderSelect : a Cloud Broker Framework for Cloud Renderfarm Services

Hollywood movies like the "Flight", "Star Trek: Into Darkness" showcase the magical photorealistic effects produced using the 3D animation techniques. In the 3D studios the animation scene files undergo a process called as rendering, where the 3D wire frame models are converted into 3D photorealistic images. As the rendering process is both a computationally intensive and a time consuming task, the cloud services based rendering in cloud render farms is gaining popularity among the animators. The advantages of using the cloud services based render farms are that it enables the on demand scalability of the render nodes on the Pay-as-you-go model. The animators could choose from either an IaaS cloud service that provides only the render nodes or a PaaS Render farm service that provides the complete rendering environment which includes the render nodes, software licensing, render job management software etc. Though cloud render farms offer many benefits, the animators hesitate to move from their traditional offline rendering to cloud services based render farms as they lack the knowledge, expertise and the time to compare the render farm service providers based on the Quality of Service (QoS) offered by them, negotiate the QoS and monitor whether the agreed upon QoS is actually offered by the renderfarm service providers. In this paper we propose a Cloud Service Broker (CSB) framework called the RenderSelect that helps in the dynamic ranking, selection, negotiation and monitoring of the cloud based render farm services. The cloud services based renderfarms are ranked and selected services based on multi criteria QoS requirements. Analytical Hierarchical Process (AHP), the popular Multi Criteria Decision Making (MCDM) method is used for ranking and selecting the cloud services based renderfarms. The AHP method of ranking is illustrated in detail with an example. It could be verified that AHP method ranks the cloud services effectively with less time and complexity.

*Keywords*: 3D Animation Rendering, Cloud Render farms, Cloud Broker Services, Ranking Cloud Services, Cloud Services Selection, Quality of Service (QoS), Decision Making, Multi Criteria Decision Making (MCDM) and Analytical Hierarchical Process (AHP).

## 1. Introduction

The process of rendering an animation file is a computationally expensive task. As the individual frames of a scene file can be processed in parallel, the concept of rendering in cloud based Render Farms is gaining popularity among the animation studios. The advantages of using the cloud based render farms are that it enables on demand scalability of the render nodes based on the pay-as-you-go model. This enable the animation studios and the freelancers to convert the Capital Expenses to Operational Expenses and increase their profit as they need not invest huge amount in buying expensive hardware and software licenses. The cloud based render farms services are of both the IaaS (Infra Structure as a Service) and PaaS (Platfrom -as-a-Service) types. The IaaS type of render farm services offer only the render nodes whereas PaaS provides not only the infrastructure but also the software required for rendering the images and the software license fee is included in the render node cost. Many of the cloud based render farm services are based on the PaaS (Platfrom -as-a-Service) delivery model. In spite of the advantages offered by the cloud render farm services, the animators and the 3d studios face many challenges in using these services.

In the animation field, the animators in the 3d studios have high level of competency and expertise in their core area of animation but may lack in-depth knowledge and expertise about using the cloud computing technology to its highest potential. Usually the animators interested in using the cloud renderfarm services use the websites of the service providers as the first point of information. But most of the service providers (SP's) have not published enough information in their websites to enable the user to check whether his functional and Quality of the service requirements will be met by the SP. Though some of the information may be available, again the animators lack the knowledge about the Quality of Service parameters that are to be considered when comparing the service obtained for a specific cost. Since there is no specific protocol followed for designing the services model offered in cloud, even comparing the services based on a single parameter such as cost is a challenge. The animators also lack knowledge about the importance of the Service Level Agreements and monitoring the services. Most of the service providers of cloud render farms sign only a Non Disclosure Agreement (NDA) and do not publish details about the SLA's in their websites. The animators also lack the time and facility to negotiate and monitor the services offered to them. These challenges faced by the animators in using the cloud based render farms indicates the need of a cloud service broker who could facilitate the various transactions between an animator and the cloud based render farm service providers.

Many works have been done on the concept of a cloud broker service for cloud computing [10] – [18]. However the concept of a cloud broker framework for the cloud render farm services is not dealt with as far as our knowledge and research is considered. The concept of a CSB for the cloud broker render farm differs from the above works as these render farm services are intended to a specific domain. Also, the functional and the non functional requirements and preferences are specific to the rendering process. For example, many of these works are focused only on the IaaS type of cloud service and CPU type of compute unit, however when coming to animation rendering in cloud, the PaaS type of cloud service and the GPU rendering is more popular among the animators and 3d studios. Thus a customized Cloud broker service that focuses on the requirements of the animation industry would be more beneficial to the animator's community who are experts in animation but a common man when dealing with cloud services.

Ruby et al [38] have explored the layer architecture for the cloud render farms but the detailed architecture was not dealt. In this paper we introduce an initial version of our Cloud Service Broker (CSB) framework called the RenderSelect for cloud based render farm service ranking, selection, negotiation and monitoring. The Contributions of this paper are 1) To provide a Cloud Broker Service frame work called the RenderSelect for the ranking and selection, negotiation and monitoring of the cloud render farm services. 2) To provide a common platform to enable the collection of the functional and the non functional requirements of the end user and the offerings from the Service Providers. 3) A render farm service selection algorithm to enable the animators to indentify the cloud render farm service that satisfy their functional and Quality of service requirements. 4) Solve the render farm service selection as a Multi Criteria Decision Making (MCDM) problem using the Analytical Hierarchical Process (AHP) method and rank the services. In this paper, the architecture framework, Service Selection algorithm and the AHP method of ranking the render farm services are discussed in detail. However the SLA negotiation and service monitoring processes are not given in detail due to the lack of space and need for more research.

The rest of the paper is organized as follows: The background and the related work are discussed in the next section. Section 3 gives an overview of the RenderSelect CSB framework and explains the Service Selection Algorithm used by the RenderSelect CSB. The section 4 explains the step by step process involved in the Analytical Hierarchical Process with an illustrative example. In the section 5 a brief discussion on the results is provided. Finally the section 6 concludes the topic with the proposed future work.

## 2. BACKGROUND AND RELATED WORK

### 2.1 Renderfarms

A render farm is nothing but a cluster of computers connected together in a network for the purpose of rendering the images faster [19], [20]. Each computer in a Render farm is called a Render node. Parallel computing enables each node in a Render farm to render the specific image file submitted to it. Render management software or a Queue manager is used to automatically distribute tasks to each render node. The Render management software assesses the needs of the rendering job like quality of the image, capability of each node, current networking status etc., and assigns the job to an appropriate render node. Once a render node completes the rendering job assigned to it, another task is assigned immediately to the node and it is kept busy [19].

### 2.2 Cloud based Render farms

The other distributed computing technologies like the grid [1-5] and the cloud have also been explored for the rendering purpose and had been proved fruitful [6-9]. Cloud based rendering is similar to offline rendering, except that, in Cloud based rendering, the rendering process is done on the machines in the service provider render farms [19]. In the Cloud based rendering, the animators upload and submit their animation files to be rendered to the Cloud rendering service providers server through their web interface. The Cloud rendering service providers usually have their own rendering job queue management software on the server. The Render management software assesses the needs of the rendering job just like in the case of an offline Renderfarm and assigns the job to an appropriate render node and keeps it busy. The updates about the rendering process are displayed in the dashboard of the Render management software. The user has the privilege to manage the rendering process by stopping or pause it. The advantages of Cloud based rendering is that the rendering time can be reduced drastically by increasing the no of resources utilized for rendering on-demand and pay only for the resources utilized on an hourly basis. The Process of rendering in a Cloud based Renderfarm is given below in figure 1.

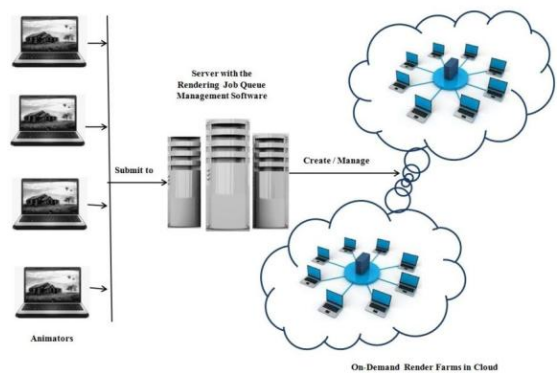

**Figure 1 Cloud based Rendering**

### 2.3 Types of cloud based Renderfarms

The three types of service models of cloud computing in general are: a) Infrastructure-as-a-Service (IaaS), Software-as-a-Service (SaaS) and Platform-as-a-Service (PaaS) [21], [22]. However the cloud based renderfarms are delivered as IaaS or PaaS models.

The SaaS is included in the PaaS model of services to provide a holistic rendering environment to the users. The IaaS type of cloud services offer the compute units as virtual machines which can be used as render nodes. However the software license requirements have to be provided and taken care of by the user. Examples of some popular IaaS service providers are Amazon Ec2 [23], Rackspace, GoGrid etc. The PaaS type of cloud services provides a complete environment for rendering which includes the render nodes, software licensing, job management software etc. Thus a user need not worry about purchasing a software license for the render engines like v-ray but can use the PaaS rendering services like the Rendering Fox [24], Rebusfarm [25] etc. The RederSelect CSB is designed for both IaaS and the PaaS.

## 2.4 Cloud Broker Services

According to Gartner cloud consumers need cloud service brokers (CSB) to unlock the potential of cloud services. He also predicts that the cloud services brokerage (CSBs) will mainly be in charge of the management, performance, and delivery of cloud services [26]. In the Cloud Computing Reference Model of the NIST, the Cloud Broker is identified as the actor in charge of service intermediation, service aggregation, and service arbitrage [27]. The concept of a Cloud Broker Service is not new and many research works related to the CSB have been published. OPTIMIS is cloud broker that supports both Independent Inter-Clouds (Multi-Clouds) and federation of clouds concepts [11]. The Deployment Engine (DE) and Service Optimizer (SO) enable the clients to launch and monitor the services within multiple cloud providers. However the major drawback is that the OPTIMIS agents should be deployed in the Cloud provider's data centers. The Contrail [12] broker system also supports the federation and independent Inter-Clouds. A major drawback of Contrail is that similar to OPTIMIS, Contrail also has the need to develop and maintain Contrail adapters that are specific to multiple vendors. The mOSAIC ia another popular broker system that facilitates the development and deployment of applications across multiple clouds [13]. It does not require any user involvement and provisions the applications on a set of predefined clouds automatically. Thus a predefined set of SLA's that serve as the performance indicators at the component and the application levels is crucial to control the brokering system. Another cloud broker service that is at an early stage of development is called the STRATOS [14]. In STRATOS, the CloudManager gets the requirements and the application topology from the user in a file called the Topology Descriptor File (TDF) and contacts the Broker component. The broker component, determines the multiple clouds for the optimal initial resource allocation. Based on the monitoring information received, the CloudManager and the Broker take decisions about the further provisioning of applications. RigthScale achieves the Application brokering through the alert-action mechanism, similar to the trigger-action mechanism [15]. The other broker services similar to the RightScale include EnStratus [16], Scalr [17] and Kaavo [18]. However these services differ based on the set of supported cloud vendors, in the pricing, technologies and the terminologies used.

## 2.5. Comparing, Ranking and monitoring cloud services

Ranking the services is an important task that facilitates the right service provider selection. Garg et al [28] proposed the SMICloud framework for comparing and ranking of IaaS cloud services based on the Service Measurement Index (SMI) suggested by the CSMIC (Cloud Service Measurement Index Consortium) [29]. Another notable work that enables the comparing of the cloud services is the "CloudCmp" by Li et al. [30],[31],[32]. The CloudCmp compares the offers from the different cloud providers in terms of performance and cost of Virtual Machines (VM's). Service Level Agreements (SLA's) for the cloud services is also another important topic of research and many works have been done in this regard. Another important task is to monitor the cloud services provider to check if the services offered are according to the standards agreed in the SLA's. A European project called the mOSAIC [34] delegates the monitoring of the resources utilization and the SLA negotiation to the Cloud Agency. Grant, Josh, and Adam Wood-Gaines have developed a Renderfarm monitoring system [33] which could be used to monitor the resource utilization within the render farms. All these related works mentioned above are for the cloud services in general. Though these ideas can be incorporated in the service selection of the render farms, further research works focused primarily on the selection and provisioning of the render farms is crucial as the cloud render farm services are more domain specific and different from the other cloud services. Thus it gives a lot of scope for the researchers to bring in new ideas in terms of SLA management and monitoring of cloud based render farm services.

## 3. THE RENDER SELECT CSB

### 3.1 Architecture Overview

The key components of the RenderSelect CSB are given in the figure 2. The RenderSelect CSB is a five layered architecture. The first layer of the RenderSelect is the CSB web interface that provides a user friendly GUI (Graphical user interface) for the end user and is responsible for the identity and the access management of the various actors involved like the users, Service Providers, brokers etc. The profiles of the actors are stored in the profiles database. The Requirements Analyzer in the second layer is responsible for collecting all the information essential for the service matching and ranking. The RGI (Requirements Gathering Interface) in this layer is responsible for collecting the functional and the non functional requirements of the end users in the corresponding templates. Similarly, the details about the functional and QoS offerings of the SP's are collected by the SDGI (Services Details Gathering Interface) by sending a common template to all the SP's. The information gathered from the end users and the SP's are stored in the requirements and the services catalogue registry respectively. The collection of information using the templates enables the matching and comparison of services as the templates sent to the end users and all the SP's have the same fields and units.

The FR_Match Maker (Functional Requirements match maker) in this layer is responsible for identifying and filtering the services that match the functional requirements of the client by searching the services catalogue database which contains all the details of the services that are associated with the RenderSelect CSB. The filtered list of services is sent to the Renderfarm Selector in the next layer. The Renderfarm Selector is responsible for the evaluation, ranking and selection of the filtered services based on the renderfarm Service Selection algorithm explained in the next section. It has two components namely the SMI based Evaluator and the Ranking System. The SMI based Evaluator, evaluates the services using the popular Analytic Hierarchy Process (AHP) Multi Criteria Decision Making Technique. The AHP technique enables the user to assign weights to SMI attributes to indicate the importance of the attribute in the selecting the service provider. The SMI based Evaluator calculates the AHP Weightings for all the renderfarm service providers. The ranking system ranks the services according to the AHP Weightings. The render farm Service provider with the highest AHP weighting is ranked as the first choice.

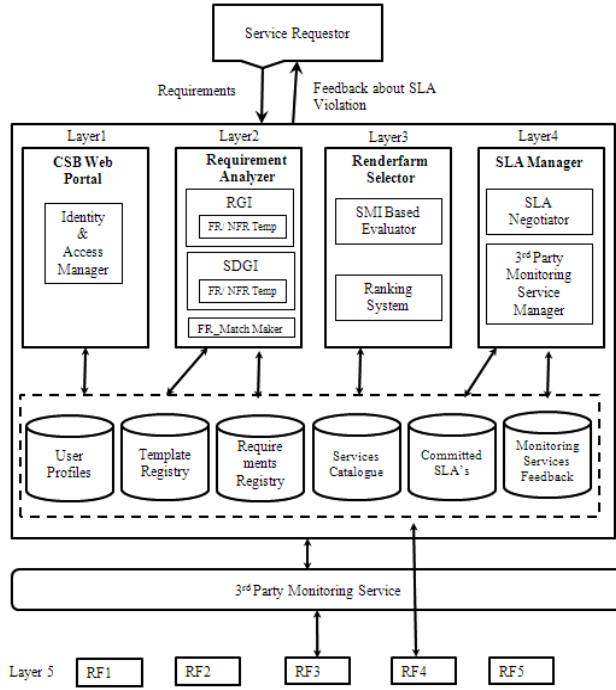

**Figure 2 Architecture diagram of the RenderSelect CSB**

The SLA Manager in the layer 4 of the CSB is responsible for the SLA negotiation and monitoring. It has two components namely the SLA Negotiator and the third party monitoring service manager. The SLA Negotiator takes care of the Service Level Agreement negotiation between the end user and the selected Renderfarm service provider. The committed SLA's after the negotiations are stored in the database called the committed SLA's database. The second component called the third party monitoring service manager connects with a third party SLA monitoring service to monitor the service that is being offered to the end user. The feedback from the third party SLA monitoring service about the SLA violations is informed to the user and is stored in the monitoring services feedback database. The service provider is penalized for the violations according to the committed SLA's. The last layer contains the GUI of the different cloud based render farm service providers whose services can be launched by clicking the GUI of the service provider selected by the user for provisioning the resources.

## 3.2. Renderfarm Service Selection Algorithm

Step 1:

RFServiceOfferings = GetRFServiceOfferings( FN_Offerings, NFN_Offerings)

Step 2:

ServiceRequirements= GetServiceRequirements(FN_Requirements, SMI_Requirements)Step 3:

FN_MatchedRFList= {RFService|FN_Matching(FN_Offerings, FN_Requirements)}

Step 4:

RankedRFList=$AHP_{rank}$(FN_MatchedRFList,SMI_Requirements)

Step 5:

Selected_RFService = Select_RFService(Max($AHP_{rank}$))

**Figure 3  Service Selection Algorithm**

The Renderfarm Service Selection algorithm is used for matching, ranking and selecting the right service provider who could satisfy both the functional and the non-functional requirements of the end user. The Renderfarm Service Selection algorithm is given below in figure 2. In the step 1, the Functional Offerings and the SMI based Non-Functional QoS offerings of the service providers are obtained using the GetRFServiceOfferings() function. The service requirements of the users are fetched using the GetServiceRequest() function in the step2. In the step 3, the Render Farm services are matched functionally using the FN_Matching() function and the functionally matched candidate services list FN_MatchedRFList is obtained. In the step 4, the matched list of the service providers (FN_MatchedRFList) is passed on to the AHP rank () function as an argument to filter the services that match both the functional and the non functional requirements of the end users and the ranked Render Farm service providers list (RankedRFList) using the AHP technique is obtained. In the final step 6, the render farm service provider with the maximum AHP ranking is selected as the best render farm service provider who could satisfy both the functional and the non functional requirements of the end user.

## 4. RANKING OF CLOUD RENDER FARM SERVICES

### 4.1  SMI Metrics for Ranking
A cloud service should have the ability to satisfy both the Functional (FN) and the Non Functional (NFN) Requirements of the client/ user. The functional requirements are essential for completing the task. However, the Non functional requirements are the quality that qualifies a good cloud render farm service but is not essential to complete the task. Some examples of the functional requirements of the cloud based render farm include the versions of the software that are supported like the 3ds max 2009, Maya 7.0 etc, the render engines supported like the Mental Ray, V-Ray etc, the Render node configuration ( Memory, CPU, disk space etc). The non functional requirements include the SMI metric attributes like the service response time, availability, elasticity etc. The CSMIC (Cloud Service Measurement Index Consortium) has suggested the Service Measurement Index (SMI) for comparing the cloud based services effectively [29]. The top level SMI metrics considered by the RenderSelect CSB  are the Accountability, Agility, Assurance, Cost, Performance and Security.

### 4.2  MCDM and Analytical Hierarchy Process(AHP)
The problem of selecting a service by evaluating multiple criteria is termed as a Multi Criteria Decision making problem and many methods like the analytical Hierarchical Process (AHP), Goal Programming etc are available in the domain of operations research to solve the MCDM problems. The AHP technique used in this work is  very popular among the others as it is a very simple and an effective method . It is used in the decision making situations which involves selection of one QoS property from several other alternatives on the basis of multiple criteria with competing nature.

## 4.3 The Analytical Hierarchical Process (AHP) method

The step by step ranking procedure of AHP for ranking five cloud based render farm services with 7 Qos attributes is explained in detail in this section. The AHP hierarchy enables the decision maker to specify the overall goal, criteria, sub-criteria and the decision alternatives in the form of a hierarchical diagram. The AHP hierarchical diagram for the selection of the render farm service provider is given below in the figure 4.

*Step: 1 Decompose the Render farm selection problem:*

The underlying Render farm service selection problem (multi-criteria decision-making problem) is decomposed according to its main components. In the AHP hierarchy diagram for selection of cloud render farm service provider, the first layer specifies the overall goal, which is to estimate the relative Service Management Index (SMI) of the services, rank the services and select the best cloud render farm service from the ranked list of services that satisfy the user QoS requirements.

*Step: 2 Define the criteria for cloud Render farm service selection:*

The criteria for cloud Render farm service selection is expressed as the Required QoS group of attributes and the Optional QoS attributes group in the second layer. The second layer of the AHP hierarchy diagram represents these criteria. At this level the individual QoS attributes are grouped together as the Required QoS attributes ( QR ) with weight (WR) and the Optional QoS attributes (QO) group with weight (WO) .

*Step: 3 Design the QoS attributes hierarchy for QoS Ranking:*

To design the hierarchy of the QoS attributes at the Top level and at the sub levels, the desired SMI metrics like the Accountability, Agility, Assurance, Cost and Performance are placed as the Top level metrics in the third layer and the sub-criteria or the Sub – level attributes of each of the top level SMI metrics is placed in the fourth layer. The sub level metrics considered here are: Service Stability, adaptability, Elasticity, Availability, Service Stability (Upload time), Render node cost, Service Response Time. The last layer consists of the various renderfarm service alternatives for selection.

### 4.3.1 Perform Pairwise Comparison and Prioritization

The next step in the AHP is to perform pair wise comparison and prioritization of the attributes. It starts from the lower level sub-attributes to the Top level SMI attributes. In order to perform the pair wise comparison of the cloud render farm services, the relative importance of each QoS attribute and group within each level has to be estimated. The relative importance of each attribute and group can be calculated by assigning weights to each QoS attribute and group within each level. In our work we use the relative weighting method and thus there is no need to normalize the SMI attribute values. In the Relative weighting method the weights are assigned universally to each of the QoS attribute in a group and the criteria is that the sum of all the relative weights of QoS attributes in each group should be equal to one.

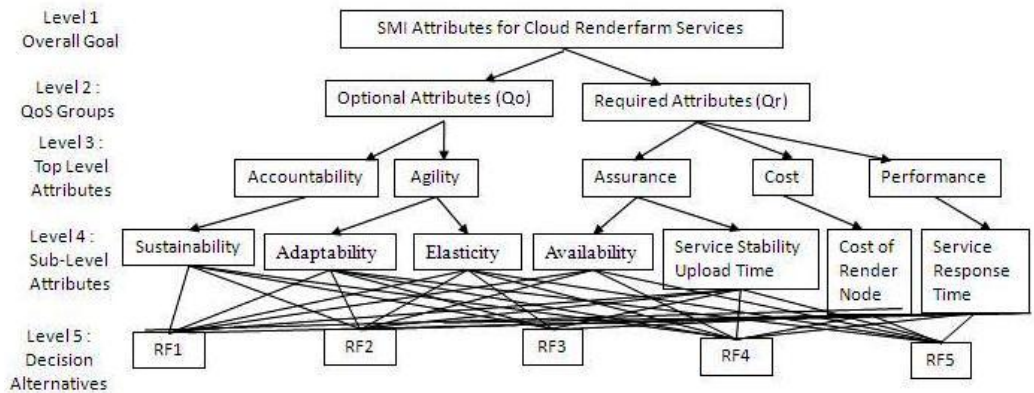

**Figure : 4 AHP hierarchy for selection of cloud render farm service provider**

### 4.3.1.1 Assumptions for pairwise comparison of SMI attributes

Let $W_i$ be the weight assigned by the user for the SMI attribute $a$ , $V_m$ and $V_n$ be the values of the attribute $a$ for the renderfarm service $m$ and $n$.
Let $R_i$ and $R_j$ be the cloud renderfarm services and the relative ranking of the renderfarm $R_i$ over $R_j$ is indicated by $R_i / R_j$.
Also let $V_{Req}$ be the required QoS value of the service specified by the user.

### 4.3.1.2 Rules for comparing the attribute values

#### 4.3.1.2.1 QoS value type and its unit

The important criterion to be considered during the comparison of the values is the QoS value type and its units. The QoS value type of an attribute may be a single, multiple, Fuzzy, simple or complex type. The rules to be followed for comparing the values of different value types is discussed in detail by Tran et al in [35].In the RenderSelect CSB, since the same SMI template is used to collect the information from both the user and the service provider, the set of attributes compared have the same data type and units.

#### 4.3.1.2 QoS Tendency

The QoS tendency can be considered as the impact direction of an attribute. The QoS tendency of an attribute may be positive, negative, close or exact. For a QoS attribute with a positive tendency, the higher the value of the attributes the better the renderfarm services offered. Hence, the Renderfarm Relative Rank Matrix (*RRRMmn*) of the service *Sm* over the Service *Sn* for a particular attribute *a* is given by,

$$RRRMmn = Sm / Sn \quad (1)$$

And for all Qos property a with a positive tendency, we have,

$$Sm / Sn = 1/ Sn / Sm \quad (2)$$

If the QoS attribute has a negative tendency, then the lower value is considered better and the Renderfarm Relative Rank Matrix (*RRRMmn*) of the service *Sm* over the Service *Sn* for a particular attribute *a* is given by,
*RRRMmn* = *Sn / Sm* (3)
In case, the QoS values of both the renderfarm services are different from the value requested by the end user, the service with the closer value is selected as the better option and if one of the service has the exact value as specified by the end user then that particular service is evaluated as the better option. More details on rules for evaluating different types of data can be found in the work of Tran et al [35].

#### 4.3.1.2 QoS Grouping

QoS Grouping enables the user to make a better choice of the service preferred. In this work, the SMI attributes are grouped into two namely the Required QoS attributes and the Optional QoS attributes. The main advantage of allowing the attributes to be grouped as 'Optional' is that, they enable the user in selecting the Renderfarm services that may not satisfy some optional QoS criteria but have a high value for the QoS attributes actually required by the user and hence a better choice of cloud renderfarm service that satisfy the user requirements can be identified effectively.

In this work, the groups $Q_O$ and $Q_R$ are considered to have the weights $W_R$ and $W_O$ assigned to them by the user and since the relative weighting is used,it is assumed that the sum of weights is one ($W_R + W_O = 1$). Usually the value of the weights $W_R$ is set higher then the value of $W_O$ by the user. For example, $W_R = 0.6$ and $W_O = 0.4$ indicates that the Required QoS properties are more important than the Optional QoS group. The details of the SMI attributes , groups , their Qos values, weights etc are given in the table 1 given below. In our work we use the relative weighting method and thus there is no need to normalize the SMI attribute values.

Table.1. Numerical example values of SMI attributes for ranking cloud Renderfarm services using AHP.

| QoS Group Level | $Q_O$ (Optional Attributes) ($W_O$=0.4) | | | $Q_R$ (Required Attributes) ($W_R$=0.6) | | | |
|---|---|---|---|---|---|---|---|
| Top Level Attributes | Accountability (Wt=0.3) | Agility (Wt=0.7) | | Assurance (Wt=0.3) | | Cost (Wt=0.4) | Performance (Wt=0.3) |
| Sub - Level Attributes | Sustainability (Wt=1) | Adaptability (Wt=0.3) | Elasticity (Wt=0.7) | Availability (Wt=0.6) | Stability (Wt=0.4) | Render Node Cost (Wt=0.4) | SRT (Wt=1) |
| Unit | Unitless | Average Time (Sec) | Average Time (Sec) | Percentage (%) | Time (Sec) | $ Per Core Hour | Average Time (Sec) |
| Value Type | Numeric | Numeric | Numeric | Numeric | Numeric | Numeric | Numeric |
| QoS Tendency | Positive | Negative | Negative | Positive | Negative | Negative | Negative |
| Requested QoS Values | >1 | 50 | 65 | 99% | 15 | <1$ | 40 |
| RenderFarm (RF1) | 1.2 | 40 | 45 | 99.90% | 10 | 0.70 | 40 |
| RenderFarm (RF2) | 1.5 | 45 | 65 | 99.99% | 12 | 0.60 | 30 |
| RenderFarm (RF3) | 1.7 | 30 | 70 | 99.90% | 15 | 0.08 | 35 |
| RenderFarm (RF4) | 1 | 50 | 35 | 99.99% | 20 | 0.10 | 25 |
| RenderFarm (RF5) | 1.3 | 50 | 40 | 99% | 10 | 0.30 | 30 |

*Step:5 Compute Renderfarm Ralative Ranking for the Sub-level attribute*

The one-to-one comparison of each cloud renderfarm service for a particular attribute could be done by forming a Renderfarm Ralative Ranking Matrix ( $RRRM_{mn}$ ) of size NR x NR Where, NR is the total number of Renderfarm Services to be compared for ranking. We start the computation of the render farm services ranking by first computing the Renderfarm Relative Ranking Matrix for upload time ( $RRRM_{UploadTime}$) as shown in the figure 5 given below:

|      | RF1   | RF2   | RF3   | RF4   | RF5   |
|------|-------|-------|-------|-------|-------|
| RF1  | 1     | 12/10 | 15/10 | 20/10 | 10/10 |
| RF2  | 10/12 | 1     | 15/12 | 20/12 | 10/12 |
| RF3  | 10/15 | 12/15 | 1     | 20/15 | 10/15 |
| RF4  | 10/20 | 12/20 | 15/20 | 1     | 10/20 |
| RF5  | 10/10 | 12/10 | 15/10 | 20/10 | 1     |

Figure 5 Renderfarm Ralative Ranking Matrix ( $RRRM_{mn}$ )

The Eigen value of the matrix also called as the Renderfarm Relative Ranking Vector ($RRRV_{mn}$) that gives the relative ranking of all the cloud renderfarm services for a particular sub-level attribute is estimated according to the step by step instruction given in the work of sathy et al [36] and Karllson[37]

Using the matrix given above, the renderfarm relative ranking vector for the Upload Time attribute $RRRV_{UploadTime}$ is calculated to be:

$RRRV_{UploadTime}$ = [0.2590, 0.2158, 0.1366, 0.1295, 0.2590 ]. The RRRV values for all such attributes is given in Table 2.

Table 2. The Renderfarm Relative Ranking vector (RRRV) value for the Sub-level SMI attributes.

| RF | Sub-level SMI Attributes | | | | | | |
|----|---|---|---|---|---|---|---|
|    | Accountability | Agility | | Assurance | | Cost Effec | Performance |
|    | Sustainability $RRRV_{SUS}$ | Adaptability $RRRV_{ADAP}$ | Elasticity $RRRV_{ELA}$ | Availability $RRRV_{AVA}$ | UploadTime $RRRV_{UPLOAD}$ | NodeCost $RRRV_{COST}$ | SRT $RRRV_{SRT}$ |
| RF1 | 0.1791 | 0.2074 | 0.2107 | 0.2003 | 0.2590 | 0.1123 | 0.1560 |
| RF2 | 0.2239 | 0.1843 | 0.1459 | 0.2006 | 0.2158 | 0.3687 | 0.2080 |
| RF3 | 0.2537 | 0.2765 | 0.1355 | 0.2004 | 0.1366 | 0.2639 | 0.1783 |
| RF4 | 0.1492 | 0.1659 | 0.2709 | 0.2006 | 0.1295 | 0.2319 | 0.2493 |
| RF5 | 0.0376 | 0.1659 | 0.2370 | 0.1985 | 0.2590 | 0.0738 | 0.2080 |

*Step: 6 Aggregate the Renderfarm Relative Ranking for the Top level attributes and QoS Groups*

Once the ($RRRV_{mn}$) for each individual attribute at the sub-level is estimated, we compute the relative ranking vector for each group of the Top level by multiplying the $RRRM_{mn}$ with the corresponding weight assigned to the sub-level attribute at level 4.

For example in order to calculate the ($GRRRV_{Assur}$) of the Assurance Top level attribute, the $RRRV_{mn}$ of the availability and the UploadTime sub–level attributes associated with it are aggregated with their weights at the level 4 as given in the equation (4). The group renderfarm relative ranking matrix (GRRRV) for all the top level attributes are given in the Table3.

$$GRRRV_{Assur} = [RRRV_{Avail}\ RRRV_{UTime}] * [W_{UTime} * W_{Avail}] \quad (4).$$

Table 3. GRRV value for Top level attributes groups

| RF | Top Level SMI Attributes Groups | | | | |
|---|---|---|---|---|---|
| | $GRRRV_{ACC}$ | $GRRRV_{AG}$ | $GRRRV_{ASS}$ | $GRRRV_{COST}$ | $GRRRV_{PERF}$ |
| RF1 | 0.1791 | 0.2097 | 0.2238 | 0.1123 | 0.1560 |
| RF2 | 0.2239 | 0.1574 | 0.2067 | 0.3687 | 0.2080 |
| RF3 | 0.2537 | 0.1778 | 0.1749 | 0.2639 | 0.1783 |
| RF4 | 0.1492 | 0.2394 | 0.1722 | 0.2319 | 0.2496 |
| RF5 | 0.0376 | 0.2157 | 0.2227 | 0.0739 | 0.2080 |

Table. 4. Aggregated Relative Ranking Vector value for groups $Q_O$ and $Q_R$

| RF | OptionalAttribute($Q_O$) | RequiredAttributes($Q_R$) |
|---|---|---|
| RF1 | 0.2005 | 0.1589 |
| RF2 | 0.1774 | 0.2719 |
| RF3 | 0.2006 | 0.2115 |
| RF4 | 0.2506 | 0.2193 |
| RF5 | 0.1623 | 0.1601 |

Then the Aggregated Group Renderfarm Relative Ranking Vector ($GRRRV_{mn}$) of the groups $Q_O$ and $Q_R$ is calculated by multiplying the value of the Top level SMI attributes ($GRRRM_{mn}$) with its corresponding weight at level 3 using the formulae (5) as given below in the table 4.

$$GRRRV_{Qo} = [GRRRV_{Assur}\ GRRRV_{Agil}] * [W_{Assur} * W_{Agil}] \quad (5)$$

*Step 7: Compute the Final relative ranking vector of the groups*
The group relative ranking vectors of the two groups *QO* and *QR* area aggregated with their corresponding weights for estimating the final ranking vector (*FRRRVmn*) as given below:

$$FRRRV_{mn} = GRRRV_{Qo} * W_O + GRRRV_{QR} * W_R \quad (6)$$

*Step 8: Sorting the list of ranked services*
In the final step, the *FRRRV$_{mn}$* is sorted to get a list of the ranked cloud renderfarm services for selection. Usually the cloud renderfarm service with the highest FRRRVmn is selected as the best service for the end user. The sorted list of final ranking vector (*FRRRV$_{mn}$*) is given below in table 5.

Table . 5. Final Overall AHP Ranking

| RF  | FinalRank | DecisionPreferences |
|-----|-----------|---------------------|
| RF1 | 0.1755    | Fourth Choice       |
| RF2 | 0.2341    | First Choice        |
| RF3 | 0.2071    | Third Choice        |
| RF4 | 0.2318    | Second Choice       |
| RF5 | 0.1610    | Fifth Choice        |

## 5. RESULTS AND DISCUSSION

The final overall AHP ranking of the renderfarm services is given in the table 5. The renderfarm service (RF2) which has the maximum (FRRRV) value is considered as the best choice for the weights assigned to the attributes. The next best option is the RF4 followed by the RF3, RF1 and RF5.

By plotting the group relative ranking of the top level attribute given in the toble 3 in the kiviat graph, useful insights about the QoS properties of the renderfarm services can be gained. The kiviat graphs of all the renderfarms is given in the figure (6-10) below.

Computing the kiviat graph of the (GRRV) of toplevel attributes with that of the final services ranking based on AHP. It is interesting to note that while the best option based on AHP ranking for the group of optional and required attributes is considered to be RF2. The kiviat graph shows that when the individual of top level attributes like performance and cost effectiveness alone is considered, the RF4 outperforms the RF2. However when the priority is on a group of optional and required attributes the required attributes, the RF2 remains as the best option.

**Kivait Graphs Of Top Level Attributes**

RRRV of RF1          RRRV of RF2          RRRV of RF3

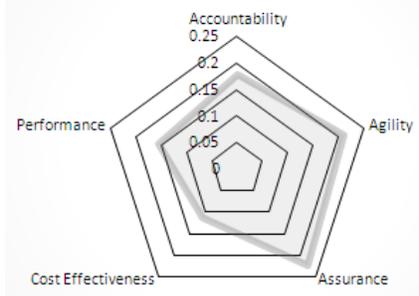 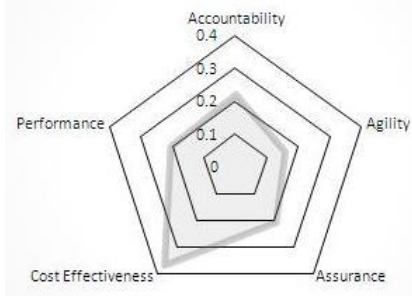 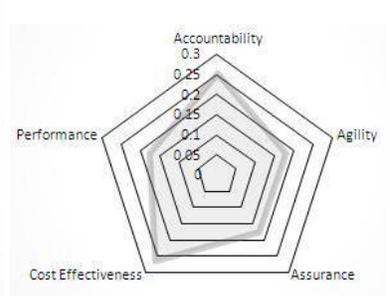

**Figure 6 Relative ranking of QoS values of RF1    Figure 7 Relative ranking of QoS values of RF2    Figure 8 Relative ranking of QoS values of RF3**

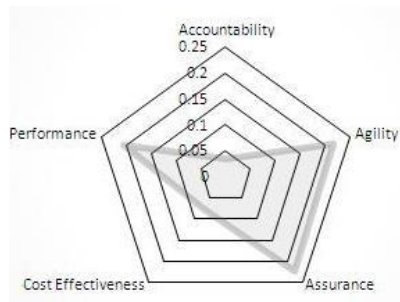
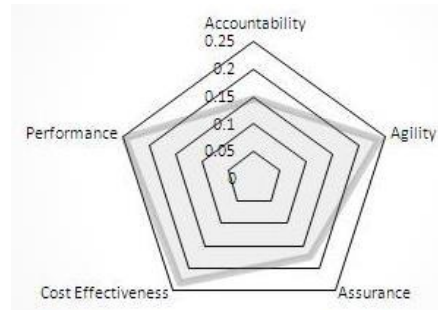

Figure 9 Relative ranking of QoS values of RF4    Figure 10 Relative ranking of QoS values of RF5

## 6. CONCLUSION AND FUTURE WORK

## 7. REFERENCES

[1] A. Chong, A. Sourin, and K. Levinski. Grid-based computer animation rendering. InGRAPHITE '06: Proceedings of the 4th international conference on Computer graphics and interactive techniques in Australasia and SoutheastAsia, New York, NY, USA, 2006. ACM.

[2] Glez-Morcillo, C., et al. "A New Approach to Grid Computing for Distributed Rendering." *P2P, Parallel, Grid, Cloud and Internet Computing (3PGCIC), 2011 International Conference on*. IEEE, 2011.

[3] Patoli, M. Z., et al. "An open source grid based render farm for blender 3d."*Power Systems Conference and Exposition, 2009. PSCE'09. IEEE/PES*. IEEE, 2009.

[4] Patoli, Zeeshan, et al. "How to build an open source render farm based on desktop grid computing." *Wireless Networks, Information Processing and Systems*. Springer Berlin Heidelberg, 2009. 268-278.

[5] Liu, Weifeng, Bin Gong, and Yi Hu. "A large-scale rendering system based on hadoop." *Pervasive Computing and Applications (ICPCA), 2011 6th International Conference on*. IEEE, 2011.

[6] James Kennedy, Philip Healy.A method of provisioning a Cloud-based renderfarm.
EP2538328 A1 [Patent]

[7] ] James Kennedy, David Healy.A method of rendering a scene file in a Cloud-based render farm. EP2538330 A1[Patent].

[8 ] Baharon, Mohd Rizuan, et al. "Secure rendering process in cloud computing."*Privacy, Security and  Trust (PST), 2013 Eleventh Annual International Conference on*. IEEE, 2013.

[9]Choi, E-Jung, and Seoksoo Kim. "Fusion Render Cloud System for 3D Contents Using a Super Computer." *Signal Processing and Multimedia*. Springer Berlin Heidelberg, 2010. 204-211.

[10] Grozev, Nikolay, and Rajkumar Buyya. "Inter-Cloud architectures and application brokering: taxonomy and
survey." *Software: Practice and Experience* (2012).
[11] Ferrer AJ, Hernndez F, Tordsson J, Elmroth E, Ali-Eldin A, Zsigri C, Sirvent R, Guitart J, Badia RM, Djemame K,et al.. OPTIMIS: A holistic approach to cloud service provisioning. Future Generation Computer Systems 2012;28(1):66–77.

[12] Carlini E, Coppola M, Dazzi P, Ricci L, Righetti G. Cloud Federations in Contrail. Proceedings of Euro-Par 2011:Parallel Processing Workshops, vol. 7155, Alexander Mea (ed.). Springer Berlin / Heidelberg: Bordeaux, France,2012; 159–168.

[13] Lucas Simarro J, Moreno-Vozmediano R, Montero R, Llorente I. Dynamic Placement of Virtual Machines for Cost

Optimization in Multi-Cloud Environments. Proceedings of the International Conference on High Performance Computing and Simulation (HPCS 2011), IEEE: Istanbul, Turkey, 2011; 1–7.

[14] Pawluk P, Simmons B, Smit M, Litoiu M, Mankovski S. Introducing STRATOS: A cloud broker service. Proceedings of the IEEE International Conference on Cloud Computing (CLOUD 2012), IEEE: Honolulu, Hawaii,US, 2012.

[15] RightScale. RightScale. Jun 14 2012. URL http://www.rightscale.com/.

[16] EnStratus. EnStratus. Jun 14 2012. URL https://www.enstratus.com/.

[17]Scalr. Scalr. Jun 14 2012. URL http://scalr.net/.

[18] Kaavo. Kaavo. Jun 14 2012. URL http://www.kaavo.com/.

[19] Daniel Tal, Rendering in SketchUp: From Modeling to Presentation for Architecture, John and Wiley, Inc. 2013. pp.41,42.

[20] Crockett,T.W. 1997. An Introduction to Parallel Rendering. Parallel Computing 23, 7, 819-843.

[21] P. Mell and T. Grance, "The NIST Definition of Cloud Computing," National Institute of Standards and Technology, Special Publication 800-145, 2011.

[22]R. Bohn, J. Messina, F. Liu, and J. Tong, "NIST Cloud Computing Reference Architecture," In Proc. of the 2011 IEEE World Congress on Services (SERVICES 2011), pp. 594-596, 2011

[23] Amazon Web Services. http://www.aws.amazon.com

[24] Renderingfox. www.renderingfox.com

[25]Rebusfarm. www.rebusfarm.net/

[26]Gartner, "Gartner Says Cloud Consumers Need Brokerages to Unlock the Potential of Cloud Services,"2009. http://www.gartner.com/newsroom/id/1064712

[27] Nair, Srijith K., et al. "Towards secure cloud bursting, brokerage and aggregation." *Web Services (ECOWS), 2010 IEEE 8th European Conference on*. IEEE, 2010.

[28] S. K. Garg, S. Versteeg, and R. Buyya, "SMICloud: A framework for comparing and ranking cloud services," in Proceedings of the 4th IEEE/ACM International Conference on Utility and Cloud Computing (UCC 2011, IEEE CS Press, USA), Melbourne, Australia, December 5-7 2011

[29] Siegel, Jane, and Jeff Perdue. "Cloud services measures for global use: the Service Measurement Index (SMI)." *SRII Global Conference (SRII), 2012 Annual*. IEEE, 2012.

[30] A. Li, et al., "CloudCmp: comparing public Cloud providers," In Proc. of the 10th annual conference on Internet measurement, pp. 1-14, 2010.

[31]A.Li, et al., "CloudCmp: shopping for a Cloud made easy," USENIX HotCloud, 2010.

[32]A. Li, et al., "Comparing public-Cloud providers," Internet Computing, IEEE, Vol. 15, pp. 50-53,2011

[33]Grant, Josh, and Adam Wood-Gaines. "Renderfarm monitoring system." U.S. Patent No. 7,577,955. 18 Aug. 2009.
[34] mOSAIC (www.mosaic-cloud.eu).

[35] Tran, Vuong Xuan, Hidekazu Tsuji, and Ryosuke Masuda. "A new QoS ontology and its QoS-based ranking algorithm for Web services." *Simulation Modelling Practice and Theory* 17.8 (2009): 1378-1398.

[36] Saaty, Thomas L. "How to make a decision: the analytic hierarchy process." *European journal of operational research* 48.1 (1990): 9-26.

[37] Karlsson, Joachim, and Kevin Ryan. "A cost-value approach for prioritizing requirements." *Software, IEEE* 14.5 (1997): 67-74.

[38] Ruby Annette and Aisha Banu. W. Article: A Service Broker Model for Cloud based Render Farm Selection. International Journal of Computer Applications 96(24):11-14, June 2014. Published by Foundation of Computer Science, New York, USA.